\documentclass[aps,prb,groupedaddress,twocolumn,10pt]{revtex4-1}

\begin{document}

\newcommand{\be}{\begin{equation}}
\newcommand{\ee}{\end{equation}}
\newcommand{\bea}{\begin{eqnarray}}
\newcommand{\eea}{\end{eqnarray}}

\title{IT from QUBIT or ALL from HALL?}

\author{D. V. Khveshchenko} 
\affiliation{Department of Physics and Astronomy, 
University of North Carolina, Chapel Hill, NC 27599}

\begin{abstract}
\noindent
Generalized $1+0$-dimensional Liouvillean dynamics describing 
deformations of the Sachdev-Ye-Kitaev (SYK) model, as well as the various 
$1+1$-dimensional dilaton and Horava-Lifshitz gravity theories, can all be mapped   
onto single-particle quantum mechanics of a non-relativistic charge propagating in a 
(generally, curved) $2d$ space and subject to a (generally, non-uniform) 
magnetic field. The latter description  
provides a standard playground for the phenomenon of Quantum Hall Effect (QHE),
thereby elucidating the intrinsically topological nature of pertinent gravity theories
and demystifying their (pseudo)holographic connection to a broad class of the SYK-like models.
\noindent
\end{abstract}
\maketitle

{\it Holographic mirages}\\

In light of the slower-than-desired progress in understanding of the great many  
quantum many-body systems  there has long been a dire 
need for finding a universal geometric (or, possibly, hydrodynamic) 
description of interacting quantum 
matter in terms of some semi-classical collective field variables. 

Historically, this idea was first implemented in the framework of classical kinetic theory formulated 
in terms of the Wigner distribution function and its moments, which description could then be further promoted to the (formally exact) phase space path integral over the corresponding field variable. 
Conceptually, such a construction can be classified as Kirillov-Kostant co-cycle  
quantization on the orbits of a given system's dynamical symmetry group. 

However, the intrinsic complexity of working with such exact, yet often intractable, formalism
brought out a variety of approximate techniques, of which the best known one is 
(non-linear) bosonization by which some aspects of the quantum dynamics of interacting (fermionic)
matter would be described in terms of shape fluctuations of the underlying Fermi surface \cite{bos}.

Albeit being quite different in its appearance, the more recent conjecture of holographic duality has been pursuing a somewhat similar goal. In this revolutionary proposal, the equivalent bosonic variables would be assumed to organize into multiplets reminiscent of the metric, vector, and scalar fields in one higher dimension
and governed by some local Einstein-Maxwell-scalar type of action (see \cite{ads} and references therein).  
 
Although vigorous attempts to put the general holographic conjecture on a solid ground have been continuing for over two decades, a satisfactory proof still remains elusive. 
This fact notwithstanding and putting the general burden of proof aside, 
much of the massive effort exercised under the auspices of the so-called   
AdS/CMT (Anti-de-Sitter/Condensed Matter Theory) branch of applied 
holography has been devoted to the heuristic 'bottom-up' approach.  

The latter offers a seemingly appealing resolution of the proverbial 'goals and means' dilemma by unequivocally postulating validity of the holographic conjecture in its broadest interpretation. As to the justification, it has been resorting to the steadfast declarations (bordering, at times, on either collective delusion or cargo cult) of its asserted success in explaining the data
on a host of (allegedly) strongly coupled condensed matter systems \cite{ads}.   
  
Specifically, in great many \cite{holo}  of the remarkably verbose and 
look-alike works (even prior to the advent of ChatGPT) on AdS/CMT 
the choice of a dual gravity theory 
and its bulk metric would be made merely on the basis of technical convenience, such as the known classical solutions and normal modes of their fluctuations, availability of computational software developed in classical general relativity, etc. 
 
However, the custom of flatly claiming applicability of the  holographic ideas to a cornucopia of condensed matter - i.e., neither supersymmetric (SUSY), or Lorentz-, or translationally- and/or rotationally-invariant (contrary to the parent Maldacena's conjecture where all these symmetries would be necessarily present \cite{ads}) - systems of $N\sim 1$ 'flavors' (for which values of $N$ the dual gravity theory would have been anything but local, contrary to the common AdS/CMT assumption) 
of strongly (albeit, for the most part, not 'very strongly', as
required for a classical treatment of the bulk gravity) 
interacting species has been showing gradual - yet, apparent - demise, as of lately \cite{holo}.  
Arguably, the opportunistic 'anything goes' approach has run out of steam
and the field has finally come to the point where it would need to be pursued differently
by seeking a more solid foundation. 

To that end, while being far more scientifically sound 
the attempts \cite{rg} to construct a general holographic 
picture from the first principles of quantum information 
(e.g., tensor networks) implementing the so-called 'IT from QUBIT' concept \cite{ads}, 
so far, have not proceeded beyond the exploratory level. 
In most cases where the bulk metric was definitively 
ascertained it was found to be of the basic $AdS$ or Lifshitz 
kind, thus casting doubts on the possibility of constructing anything 
as even nearly exotic as, e.g., the 'helical Bianchi $VII$' geometry   
that was repeatedly invoked in the AdS/CMT scenarios of the so-called 'strange metallic'   
normal state of cuprates \cite{bianchi}. 

Likewise, the previous attempts  \cite{rg} to derive holography directly from the scale-dependent renormalization group (RG) flow, thus implementing the holographic 'RG=GR' principle \cite{ads}, 
have been largely inconclusive and, thus far, produced either 
the same plain $AdS$ or, else, unrecognizable bulk geometries. 

Also, the attempts \cite{super} to establish a holographic correspondence between the bulk
AdS gravity (in the Lorentz signature) 
and an ordinary superconductor (even a weakly coupled BCS one)
hinge on the formal similarity between the $2d$ d'Alembertian operator acting 
in the so-called kinematic space and the mixed second derivative of a 
bi-local function (see, e.g., Eq.(19) below), thus falling too short of providing 
a true 'derivation' of holography.  
 
As compared to all the questionable (and, for its most part, easy to debunk \cite{dvk1}) 
evidence purported to be consistent with AdS/CMT, 
the recent studies of the holography-like correspondence \cite{syk} between the 
ensemble-averaged quantum mechanical SYK model \cite{sy} in $1+0$ dimensions  and
Jackiw-Teitelboim (JT) gravity \cite{jt} in $1+1$ dimensions may seem to have finally 
delivered a strong argument supporting 
the holographic conjecture (albeit in a form which is rather different from the 
earlier 'ad hoc' AdS/CMT constructions \cite{ads}). 

At the very minimum, the following discussion aims at 
extending the list of holographically dual  
$1+0$- and $1+1$-dimensional problems  beyond the extensively studied case of SYK-JT. 
It is argued that, conceivably, this specific example may represent a 
more general equivalence between a whole class of the deformed SYK models and 
a certain family of generalized $2d$ gravities.

Even more importantly, taken at its face value the SYK-JT duality raises an important 
question as to whether or not any (or all) 
instances of actually proven - as opposed to merely assumed - 
cases of holographic correspondence would be limited to those situations where the bulk theory 
appears to be of a (possibly, implicit) topological nature? 

In the specific case of SYK-JT, the bulk system does happen  
to be intrinsically topological, akin to QHE.
Therefore, should the answer to the above question happen to be affirmative, 
it would naturally explain the otherwise rather baffling  
duality between certain systems of (ostensibly) different dimensionalities, 
as per the central holographic conjecture. 
Also, it would prompt one to look for a hidden 'Hall-ness' in any situation 
where some holographic features may have been observed. 
\\

{\it From SYK to Liouville via Schwarzian}\\

Extensions of the original SYK model are described by a generic Hamiltonian  
\be
{\hat H}=\sum_q\sum_{i_1\dots i_q}J_{i_1\dots i_q}
{\hat \chi}_{i_1}\dots{\hat \chi}_{i_q}
\ee
which combines the products of some even numbers $q$ of
the $N$-colored Majorana or Dirac fermion operators ${\hat \chi}_i(\tau)$ where $i=1,\dots, N$ \cite{sy}.
In turn, the independent Gaussian-distributed 
classical random amplitudes $J_{i_1\dots i_q}$ of the all-to-all $q$-body entanglement
are characterized by the variances 
\be
{\overline {J_{i_1\dots i_q}J_{j_1\dots j_q}}}=J^2_q\prod_{k=1}^{q}\delta_{i_k,j_k}
\ee
The analysis of the model (1) 
typically starts by integrating the fermions out, thereby  
arriving at the action in terms of the bi-local field $G(\tau_1,\tau_2)$
which represents the fermion propagator and the corresponding 
self-energy $\Sigma(\tau_1,\tau_2)$ \cite{sy} 
\bea
S[G,\Sigma]={N\over 2}\int d{\tau_1}\int d{\tau_2}
(\ln(\partial_{\tau_1}\delta(\tau_1-\tau_2)-\Sigma(\tau_1,\tau_2))\nonumber\\
+\Sigma(\tau_1,\tau_2)G(\tau_1,\tau_2))-
F[G(\tau_1,\tau_2)]~~~~~~~~~~~~~~~~
\eea
where the functional $F[G]$ results from the Gaussian averaging.
Moreover, Eq.(2) can be further promoted to a retarded and/or non-uniform disorder 
correlation function, thus introducing a notion of spatial 
dimensions and space/time-dependent (retarded and/or non-local) entanglement-like couplings \cite{dvk3}. 

Solving for the self-energy $\Sigma={\delta F/\delta G}$, 
the Schwinger-Dyson equation derived from (3) can be cast in the form 
\be
\int d{\tau}(\partial_{\tau}\delta(\tau_1-\tau)+ {\delta F\over \delta G(\tau_1,\tau)} )
G(\tau,\tau_2)=\delta(\tau_1-\tau_2)
\ee
In the original $SYK_q$ model with $F[G]=J^2G^q$ 
Eq.(4) remains invariant under the infinite group 
$Diff(S^1)$ of reparametrizations (diffeomorphisms) of the thermal circle
$\tau\to f(\tau)$ with the periodicity condition 
$f(\tau+\beta)=f(\tau)+\beta$, as long as
the derivative term is neglected and provided that $G$ and $\Sigma$ transform as 
\bea 
G_f(\tau_1,\tau_2)=(\partial_{\tau_1}f(\tau_1)\partial_{\tau_2}f(\tau_2))^\Delta G(f(\tau_1),f(\tau_2))~~~\\
\Sigma_f(\tau_1,\tau_2)=(\partial_{\tau_1}f(\tau_1)\partial_{\tau_2}f(\tau_2))^{1-\Delta}\Sigma
(f(\tau_1),f(\tau_2))
\nonumber
\eea
The above properties of Eq.(4) single out a translationally-invariant  mean-field solution
(hereafter, $\tau=\tau_1-\tau_2$ and $\beta$ is the inverse temperature) \cite{sy}
\be 
G_0(\tau_1,\tau_2)=({\pi\over \beta\sin(\pi\tau/\beta)})^{2\Delta}  
\ee
In the zero temperature limit and for $J\tau\gg 1$ it demonstrates 
a pure power-law ('conformal')
behavior $G_0(\tau_1,\tau_2)\sim 1/(J\tau)^{2\Delta}$ with the fermion dimension
$\Delta=1/q$. 

This solution spontaneously breaks the full reparametrization symmetry down to its 
three-dimensional subgroup $SL(2,R)$ implemented through the Mobius transformations 
$ 
\tau\to{(a\tau+b)/{(c\tau+d)}}
$
where $ad-bc=1$, under which the solution (6) and the action (3) remain invariant. 

The reparametrization transformations outside the $SL(2,R)$ subgroup 
modify the functional form of $G_0$, thus 
exploring the entire coset $Diff(S^1)/SL(2,R)$
and providing it with the structure of a co-adjoint Virasoro orbit. 
The deviations from (6)
are controlled by the short-time expansion 
\be
\delta G_f(\tau_1,\tau_2)={\Delta\over 6}\tau^2 
Sch {\{} f(T),T {\}}G^2_0(\tau_1,\tau_2)+\dots
\ee 
Hereafter $T=(\tau_1+\tau_2)/2$ and $Sch$ denotes the Schwarzian derivative,  
$
Sch{\{}f,x {\}}= {f^{\prime\prime\prime}/f^{\prime}}-{3\over 2}
({f^{\prime\prime}/f^{\prime}})^2
$ 
(here $f^{\prime}=df/dx$) 
which obeys the differential 'composition rule'
%\be
$Sch {\{}F(f),x{\}}=Sch {\{}F(f),f{\}}{f^\prime}^2+Sch {\{}f,x{\}}$
%\ee
. 

The dynamics of the variable $f(\tau)$ is  
then governed by the non-reparametrization invariant,
yet manifestly geometrical and $SL(2,R)$-invariant, action 
\be
S_0[f]=
-{N\over Jq^2}\int d\tau Sch{\{}\tan{\pi f\over \beta},\tau {\}}
\ee
that stems from the trace of the (infrared-irrelevant in the RG sense) 
time derivative $\partial_{\tau}G$ in the gradient expansion 
of the first term in Eq.(3). 

The mean-field ('large-$N$') SYK solution (6) 
is only applicable for $1/J\ll\tau,\beta\ll N/J$, under which conditions 
the fluctuations $\delta G$ about the saddle point $G_0$ remain small. 
By contrast, in the 'Schwarzian' (long-time, low-temperature, $N/J\lesssim\tau,\beta$) regime 
these fluctuations can grow strong, 
thereby significantly altering the mean-field behavior \cite{syk,bak}. 

Namely, upon the Langer transformation  
$ 
\partial_{\tau}f=e^{\phi}
$
the Schwarzian action () reduces to 
the (ostensibly) free expression in terms of the (unbounded) variable $\phi(\tau)$
$
S_0[\phi]\sim \int d\tau(\partial_{\tau}\phi)^2
$.
However, the true Liouvillean action remains strongly non-Gaussian,  
as follows from the analysis of the products of propagators
\bea 
<G_f(\tau_1,\tau_2)\dots G_f(\tau_{2p-1},\tau_{2p})>=\nonumber\\
=\int D\phi e^{-S_0[\phi]}
\prod_{i=1}^p
{e^{\Delta(\phi(\tau_{2i-1})+\phi(\tau_{2i}))}\over 
(\int_{\tau_{2i-1}}^{\tau_{2i}}d\tau e^{\phi})^{2\Delta} }
\eea
Computing such amplitudes requires the denominator to be promoted to the exponent 
in the form of the $2p$ consecutive quenches under the action of the 
local 'vertex' operators $e^{\Delta\phi(\tau)}$.  
The resulting effective action 
\be
S[\phi]={N\over Jq^2}\int d{\tau}({1\over 2}(\partial_{\tau}{\phi})^2+J^2e^{2\phi})
\ee 
then acquires 
the exponential $1d$ Liouville potential $V_{2,2}(\phi)=J^2e^{2\phi}$
(hereafter, $V_{a,b}(\phi)$ denotes a potential which behaves as $e^{(a/b)\phi}$ 
in the limits $\phi\to\pm\infty$, respectively). 

The action $S[\phi]$ can be further quantized by switching to the Hamiltonian picture and
substituting the momentum $\pi=\delta S/\delta\partial_{\tau}\phi$
with $-i\partial_{\phi}$.
Consequently, one arrives at the (static) eigenvalue equation   
\be
-\partial^2_{\phi}\psi+V_{2,2}(\phi)\psi=E\psi
\ee 
The spectrum of Eq.(11) is continuous, $E_k=k^2$, and consists of the eigenstates  
$\psi_k(z)\sim {\sqrt {\nu\sinh\pi\nu}}K_{i\nu}(k{z})$ where $\nu={\sqrt {E-1/4}}$ and $z=J\beta e^{\phi}$.
The exact wave functions can be used to compute the various matrix elements 
$<0|e^{\Delta\phi}|k>$ explicitly. Their calculation 
reveals a universal behavior of the averaged products of an arbitrary number of propagators   
in the long-time/low-temperature regime ($N/J\lesssim\tau,\beta$) where one finds 
$
<G^p_f(\tau,0)>\sim {1/(J\tau)^{3/2}}
$ for any $p\geq 1$ and $q\geq 2$ \cite{bak}.
This behavior is markedly different from the (non-universal)  mean-field one 
at short times/high temperatures ($1/J\ll\tau,\beta\ll N/J$),
$
<G^p_f(\tau)>\sim G^p_0(\tau)\sim 1/{J\tau}^{2p/q}
$.

The intrinsically non-Gaussian nature of the action (10)
is manifest. Indeed, in the absence of the exponential term in (10) 
the latter amplitude would have been governed by
the non-logarithmic correlator $<\phi(\tau)\phi(0)>_G\sim J\tau$, thus 
demonstrating exponential, rather than algebraic, decay,
$<G^p_f(\tau)>_G\sim G^p_0(\tau)\exp(-{1\over 2}p^2\Delta^2<\phi(\tau)\phi(0)>_G)
$ which is also non-universal as a function of $p$ and $q$. 
 
Notably, the $1d$ action (10) is for just one variable representing the fluctuations of
a single soft (energy) mode. It can be readily extended to include other degrees of freedom - as, e.g., 
in the case of the complex-valued ('Dirac', as opposed to 'Majorana') 
variant of the SYK model, the additional $U(1)$ scalar field corresponding to the charge fluctuations \cite{syk}.  
\\

{\it SYK deformations}\\

A deformation of the 'potential' part of the bi-local 
Liouvillean action (3) can generally be represented in terms of a two-time integral    
with the kernel 
\be
F[G]=\sum_n{c_n}\int d\tau_1\int d\tau_2 G^n(\tau_1,\tau_{2})
\ee
with the coefficients $c_n\sim{NJ^2_n/q^2}$
which results from the ensemble-averaged 
partition function and generalizes the original SYK model described by  
the single $n=q$ term.
Additional powers of $G$ could also 
emerge if random amplitudes of the $n$- and $m$-body terms
developed some (physically quite plausible) cross-correlations, resulting in ${\overline {J_nJ_m}}\neq 0$.

Beyond the Liouville point in the multi-dimensional Hamiltonian 
parameter space, the previous analyses of the action given by Eqs.(3) and (12) have been largely limited to 
the $SYK_q-SYK_{q/2}$ model
with only two non-zero coefficients, $c_q=J^2N/2q^2$ and $c_{q/2}=2\Gamma^2N/q^2$.
For $q=4$ it has been rather extensively discussed  in the context of random tunneling between two SYK quantum dots \cite{qdot,dvk2}, the amplitude $\Gamma$ being a variance of the tunneling amplitude. 
This action also finds its applications in theoretical cosmology ('traversable wormhole') and 
discussions of the $1+1$-dimensional analog of the Hawking-Page curve \cite{worm}.  

Moreover, in most of the previous analyses for $\Gamma\ll J$
and $\Gamma\gg J$ the terms with $n=q/2$ and $n=q$ would be treated, respectively, 
as small perturbations of one another. Specifically, for $\Gamma\ll J$
and at relatively short times, $1/J\ll\tau\ll J/\Gamma^2$, the value of the fermion dimension 
$\Delta=1/q$ would be determined by the $n=q$ term, while for $\tau\gg J/\Gamma^2$ the $n=q/2$ 
term takes over, thus causing a faster decay governed by $\Delta=2/q$. 

Such analysis can be potentially 
misleading, though, as it focuses on the soft ('angular' or 'along-the-valley')
fluctuations about a chosen 
mean-field solution, while under a perturbation the mean-field solution itself might undergo a significant
change which would then require a tedious account of the hard ('radial' or 'out-of-the-valley') fluctuations. 
It can be avoided, though, by using the proper solution of the mean-field equation derived with the use of the entire functional (12).
 
Of a particular interest are crossovers between different 
conformal fixed points where all pertinent coupling constants are of
the same order. Such 'SYK transits' are not directly amenable 
to perturbation theory in the vicinity of the fixed points in question  
but can still be explored in the large-$q$ limit (see the next Section).
To that end, one can utilize the already available -and seek out new - 
non-perturbative (in general, non-conformal) mean-field solutions  \cite{dvk4}.

By analogy with the pure Liouvillean action (10),
the canonical quantization procedure applied to the 
action $S_0+\Delta S$ given by the sum of Eqs.(8) and (12)
substitutes its non-Gaussian part with the ordinary single-time integral
$\Delta S(\phi)=\int d{\tau}V(\phi)$ where 
\be
V(\phi)=\sum_n c_n e^{2n\phi/q}
\ee
For one, the aforementioned two-term action    
with the non-zero coefficients $c_q$ and $c_{q/2}$ features the Morse potential \cite{syk,dvk4} 
\be
V_{2,1}(\phi)=c_qe^{2\phi}+c_{q/2}e^{\phi}
\ee 
For both $c_q$ and $c_{q/2}$ positive the potential (14) features  
a continuous positive definite 
spectrum, $E_{\nu}=\nu^2+\lambda^2+1/4$,
with $\lambda=c_{q/2}/c_{q}$ and the eigenstates 
\bea
\psi_{\nu}(\phi)\sim
{\sqrt {\nu\sinh(2\pi\nu)}}
\Gamma({1/2}-\lambda+i\nu)W_{\lambda,i{\nu}}(2\lambda e^\phi)
\eea
where $W_{\lambda,i\nu}$ is the Whittaker function. 
For $c_{q/2}=0$ Eq.(15) reduces to the 
aforementioned eigenstates of the Liovillean potential 
given by the modified Bessel functions.
 
By contrast, for $c_{q/2}$ negative the potential develops a minimum 
and the spectrum includes ${\cal N}=[\lambda-1/2]$ bound states at 
the negative energies $E_n=\lambda^2+1/4-(n-\lambda+1/2)^2$
where $n=0,\dots, {\cal N}$.
The corresponding eigenstates are given by the associated Laguerre polynomials
\be
\psi_{n}(\phi)\sim e^{(\lambda-n-1/2)\phi-e^{\phi}} L_n^{2\lambda-2n-1}(2e^{\phi})
\ee 
At low temperatures ($\Gamma^2\beta/J\gg 1$) the number 
$\cal N$ of bound states increases and they become nearly equidistant, as in
the harmonic oscillator potential.

Notably, for $J=2\Gamma$ the 
aforementioned monotonic and non-monotonic Morse  potentials conspire to form 
a doublet of super-partners 
%\be
$V_{\pm}(\phi)=W^2(\phi)\pm{\partial_{\phi}}W(\phi)$
%\ee
with $W(\phi)\sim e^{\phi}$.    
The ground state of the binding potential $V_{-}$   
then takes on the form $\psi_0(\phi)\sim\exp(-\int Wd\phi)$.

Conceivably, the effective action $S(\phi)$ 
may develop other interesting regimes at the points of still higher symmetry. 
Albeit being special, the integrable potentials may also provide insight 
into the general behaviors. A similar situation has long been known in the physics of integrable $1d$ 
spin chains of arbitrary on-site spin.

In particular, below we demonstrate the emergence of the Toda-like action   
(in cosmology, a.k.a. 'oscillatory tracker model') described by the classically 
solvable two-term potential 
\be  
V_{2,-2}(\phi)=c_qe^{2\phi}+c_{-q}{e^{-2\phi}}
\ee
which has only discrete levels. 
Its linearly independent solutions are given by the approximate formulas
$
\psi_{\pm}(\phi)\sim \exp(\pm e^\phi-\phi/2)
$.
Notably, both Eqs.(14) and (17) belong to the still broader family of 'quasi-solvable' potentials 
$V(\phi)=c_qe^{2\phi} + c_{q/2}e^{\phi} + c_{-q/2}e^{-\phi} + c_{-q}e^{-2\phi}$. 

In the context of the problem of tunneling between two SYK quantum dots \cite{qdot},  
going into the strong-coupling regime and taking into account   
multiple tunneling processes  
can be achieved by replacing $G$ computed to zeroth order in tunneling
with the all-order expression $G/(1+i\sigma G)$ where $\sigma$ 
is the tunneling conductance \cite{dvk2}.

The corresponding potential $V(\phi)$ can then consist of an infinite number of terms.  
In that regard, especially interesting is
the 'hypersymmetric' Hulten potential $V_{0,1}(\phi)\sim{e^{\phi}/( 1-e^{\phi})}$
 with all the coefficients being equal, $c_{nq/2}=c$ for $n\geq 1$.  
It develops the $\sim 1/\phi$ behavior at 
small $\phi$, reminiscent of the Coulomb potential.
Unlike the latter, though, it features only a finite number $[\lambda]$ of bound 
states at $E_n=-((\lambda^2-n^2)/2\lambda n)^2$. 

Another interesting ('variable scaling') model was proposed in Ref.\cite{hybrid}. 
It includes an infinite number of terms with the coefficients 
$c_{nq/2}\sim n^{\nu}$. Performing an approximate summation over $n$ one obtains a power-law potential 
$ 
V_{\infty,0}(\phi)=\sum_nn^{\nu-1}e^{n\phi}\sim {1/(-\phi)^{\nu}}
$ generalizing the Coulomb one. 
\\

{\it Large $q$ limit}\\

An alternate approach to the generalized SYK-like models and a further justification of substituting 
Eq.(13) for (12) exploits the large-$q$ approximation to the propagator \cite{syk}
\be
G(\tau_1,\tau_2)={1\over 2}sgn \tau (1+{2\over q}g(\tau_1,\tau_2)+\dots)
\ee
The higher order terms $O(1/q^2)$ can also be evaluated, albeit at increasingly prohibitive costs  \cite{syk}. 
The path integral over the field $g$
is governed by the action 
\be
S(g)={N\over q^2}
\int \tau_1\int d\tau_2 ({1\over 2}{\partial_{\tau_1}}g{\partial_{\tau_2}}g+V(g))
\ee
where the potential is given by Eq.(13) as a function of the bi-local field $g(\tau_1,\tau_2)$.

A complete theory (19) is genuinely two-dimensional, 
the relative $\tau$ and 'center-of-mass' $T$ time variables playing the roles
of the effective 'radial' and 'angular' coordinates in the $2d$ 'kinematic space', respectively \cite{ads3}. 
So it is only by focusing on the former dependence and neglecting the latter 
can one reduce the low-energy sector of (19) to the $1d$ action akin to that given by Eqs.(8) and (13).

This way, one arrives at the equation of motion
\be
\partial^2_{\tau}g(\tau)=-{\partial_g V(g(\tau))}
\ee
whose solutions correspond to the mean-field configurations,
thus yielding the mean-field propagator $G_0(\tau)=\exp(2g(\tau)/q)$.  
A solution to Eq.(20) provides one with the means to probe the system's thermodynamics.
To that end, by solving (20) 
\be
\tau=\int^{0}_{g_0} {dg\over {\sqrt {V(g_0)-V(g)}}}
\ee
and putting $\tau=\beta/2$ one can compute the mean-field energy \cite{hybrid}
\be
E={N\over 4q^2}(\beta V(g_0)-2^{3/2}\int^{0}_{g_0} dg{\sqrt {V_0-V(g)}}   )
\ee
where $g_0<0$ is the turning point of the potential $V(g)$. 

In the case of the Morse potential
(14) with $g$ substituted for $\phi$ 
the explicit saddle point solution of (20) reads \cite{hybrid}
\be
g_0(\tau)=\ln{2A\sin^2\theta\over \cos({2\omega\tau/\beta}-\omega)+\cos\theta}
\ee
where $A={\sqrt {{(\omega/\beta J)^2+(\Gamma/J)^4}}}$, $\theta=\tan^{-1}(\omega J/\beta\Gamma^2)$, and 
the $\omega$ obeys the equation
$2\omega^2=(\beta\Gamma)^2+A(\beta J)^2\cos\omega$.
For $\Gamma\ll J$ it takes the values 
$\omega=\pi/2-O(1/\beta J)$ and $\omega=\pi/2-O(\Gamma^2\beta/J)$
for $1/J\ll\beta\ll 1/\Gamma$ and $1/\Gamma\ll\beta\ll J/\Gamma^2$, respectively.
 
In the zero-temperature limit, Eq.(18) yields 
\be
G_0(\tau)={1\over 2}{sgn \tau\over (1+{\sqrt {J^2+4\Gamma^2}}\tau+\Gamma^2\tau^2)^{2/q}}
\ee
As compared to the approximate conformal propagator characterizing 
the original SYK model, this expression  
is UV-finite and naturally regularized at $\tau\lesssim min[1/J,1/\Gamma]$.
Also, in contrast with the perturbative results of Refs.\cite{bak,qdot},
the saddle-point solution (24) is applicable at all $\Gamma/J$, large and small.

Gaussian fluctuations $\delta g(\tau)$ about the saddle-point 
solution of Eq.(20) are governed by the action  
\be
\delta S={N\over 2q^2} \int{d\tau} 
((\partial_{\tau}\delta g)^2+W(g_0(\tau))\delta g^2)
\ee
featuring the potential $W(g(\tau))={\partial^2_gV(g)}=\sum_nc_nn^2e^{ng}$
which is functionally similar to $V(g)$ given by (13)
and has to be evaluated at the solution $g_0(\tau)$ of Eq.(20). 

In contrast to the Schwarzian action (10) the fluctuations are scale-invariant and their strength is independent of temperature, being instead controlled by the numerical parameter $N/q^2$
and decreasing/increasing with increasing $N$ and $q$, respectively.  
As compared to the fluctuations about the mean-field solution (6)
those associated with the one given by $g_0(\tau)$ 
correspond to the pseudo-Goldstone excitations about the fixed 'valley'
in the space of field configurations which no longer needs to be adjusted.

Another uniquely simple (and previously unexplored) situation is the case of the Toda potential 
which, upon a global anisotropic coordinate rescaling, 
reduces to $V_T(g)=J^2\cosh 2g$ and coincides with its second derivative up to a factor. 
Its classical equation of motion assumes the form of the celebrated $\sinh$-Gordon equation,   
$
{\partial^2_{\tau}g}=-J^2\sinh g
$ 
whose solution satisfying the initial condition $g(0)=0$ 
reads
\be
{g_0({\tau})}=-\ln\tan(J\tau+{\pi\over 4})
\ee
Other known (quasi)solvable potentials are likely to provide novel mean-field solutions,
alongside the associated actions for their fluctuations.   
\\

{\it Particle in magnetic field}\\

The Hamiltonians reminiscent of those discussed in the previous section
routinely arise in the problem of a non-relativistic particle subject 
to a certain $2d$ static geometry $g_{ij}(x,y)$ and a vector potential $A_{i}(x,y)$. 
By exploiting this analogy one can then replace a field-theoretical 
path integral over the fluctuating variable 
$\phi(\tau)$ with a worldline one governed by the single-particle action
\be  
S[X]=\int d{\tau}({1\over 2}g_{ij}\partial_{\tau}X^{i}\partial_{\tau}X^{j}+
\partial_{\tau}X_{i}A^{i})
\ee
where $X_{\mu}=(x,y)$.
This equivalence is limited to the contributions of all single-valued (non-self-intersecting) curves
which indeed dominate for low temperatures.

In the hyperbolic plane ($H^2$) geometry such a connection between 
 the 'particle-in-magnetic-field' (PMF) problem and the SYK model 
has been extensively utilized before \cite{sy,syk}. 
It can be further extended towards a broader class of metrics and magnetic field configurations. 
As a technical simplification one can first explore the class of diagonal bulk metrics, 
$g_{ij}(x,y)=diag [g_{xx}(x), g_{yy}(x)]$, and vector potentials
in the Landau gauge, $A_{i}(x,y)=(0,A_y(x))$, which choices facilitate 
a separation of variables in the corresponding Schroedinger equation
with the Hamiltonian
\be  
H_{PMF}={1\over 2}g^{xx}(x){\pi^2_{x}}+{1\over 2}g^{yy}(x)(\pi_y-A_y(x))^2
\ee 
where $\pi_i$ is the conjugate momentum.

For the sake of the following discussion 
the background fields can be further restricted to the power-law functions 
of the $x$-coordinate (here $l$ is a characteristic length scale akin to the '$AdS$ radius') 
\be   
g^{xx}=({x/l})^{2\alpha},~~~ g^{yy}=({x/l})^{2\beta},~~~ A_{y}=Bl({l/x})^{\gamma}
\ee
so that the interval in this (Euclidean and, in general, anisotropic) metric reads 
$
ds^2=dx^2(l/x)^{2\alpha}+dy^2(l/x)^{2\beta}
$. 

In general, the Hamiltonian dynamics described by Eq.(28) develops in the 
$4d$ phase space spanned by two pairs of canonically 
conjugated variables, $(x,\pi_x)$ 
and $(y,\pi_{y})$. However, in the chosen gauge the $y$ variable becomes cyclic and the conjugate momentum $\pi_y=k$ is conserved, as in a translationally-invariant plane-wave solution propagating along the $1d$ 
boundary of a $2d$ region. 

By comparison, the $y$ variable can be paralleled with the aforementioned 'center-of-mass' time $T$.  
In contrast, dynamics in the $x$ direction remains 
non-trivial and is analogous to the dependence on
the 'relative' time $\tau$. 

The magnetic flux through the semi-space $x\geq 0$
\be
\Phi=\int dxdy{\sqrt g}(\partial_xA_y-\partial_yA_x)=B\int 
{dxdy} ({l\over y})^{\gamma+1-\alpha-\beta}
\ee
scales with the area provided that
$
\gamma+1=\alpha+\beta
$.

A uniform magnetic field in flat space corresponds to $\alpha=\beta=0$ and
$\gamma=-1$, while its much-studied counterpart on a hyperbolic plane $H^2$ can be attained for  
$\alpha=\beta=\gamma=1$.

Quantizing the PMF  
Hamiltonian (28) and factorizing its eigenstates, 
$\Psi(x,y)=\psi(x)e^{iky}$,  
one arrives at the Schroedinger equation with the quasi-$1d$  Hamiltonian 
\be  
H={1\over 2}x^{2\alpha}{\pi^2_{x}}+{1\over 2}(x^{2\beta}k^2
-2x^{2\beta-\gamma}B\pi_x+B^2x^{2\beta-2\gamma})
\ee
which contains a triad of algebraic terms with the exponents
$2\beta,2\beta-\gamma$, and $2\beta-2\gamma$.

Moreover, the Hamiltonian (31) could acquire 
still higher powers of $x$ stemming from 
the relativistic corrections 
proportional to $(\pi_{i}-A_{i})^{2n}$ with $n>1$.

For $\alpha=1$ and with the use of the logarithmic reparametrization
$x=e^z$ one can cast Eq.(31) in the form of the ordinary
$1d$ Schroedinger equation in flat space 
with the potential $V(z)$ given by Eq.(13).
Incidentally, the metric takes the form $ds^2=dz^2+e^{-2\beta z}dy^2$.

In contrast, for $\alpha\neq 1$ the corresponding $2nd$ order 
differential equation would exhibit 
a power-law potential $V(z)=\sum_nc_nz^n$ after the reparametrization
$z=x^{1-\alpha}/(1-\alpha)$ and rescaling 
$y\to y(1-\alpha)^{-\beta/(1-\alpha)}$, 
in which coordinates the metric takes the form 
$
ds^2=dz^2+z^{-2\beta/(1-\alpha)}dy^2
$.

Moreover, for $\alpha=1$ and non-zero 
$k$ and $B$ the three-term potential in Eq.(31) reduces to only two terms, provided  that
the other two exponents are related as $\beta=0$, $\beta=\gamma$, or $\beta=\gamma/2$.

In the first two cases one obtains the Morse potential (14) with $\lambda=kl/2\gamma$ and $\lambda=Bl^2/2\gamma$, respectively. Thus, the Morse scenario extends beyond the well-known 
case of a constant field and $H^2$ space of constant negative curvature.
Nonetheless, the magnetic flux $\Phi$ 
can only be proportional to the area $\int dxdy$ for $\beta=\gamma$, but not in the other two cases.  

By contrast, the third combination of the parameters 
 yields the Toda potential (17) with $c_q=B^2l^4$ and $c_{-q}=k^2l^2$  
which conforms to the symmetric potential 
$\cosh 2z$ upon uniform re-scaling
$z\to z+{1\over 2}\ln(k/Bl)$.

For a given PMF Hamiltonian much information can be inferred 
from its resolvent 
which allows for a spectral expansion 
over its $2d$ eigenstates 
\bea
D_E(x,y| x^{\prime}, y^{\prime})=
<x,y| {1\over E-H+i0} | x^{\prime}, y^{\prime}>=\nonumber\\
=\int dke^{ik(y-y^{\prime})}\sum_{n/\nu}
{\psi_{k,\nu}(x)\psi^{*}_{k,\nu}(x^{\prime})\over E-E_{k,\nu}+i0}
\eea
where $\cosh d=1+((x-x^{\prime})^2+(y-y^{\prime})^2)/2xx^{\prime}$
and the sum/integral $\Sigma_{n/\nu}$ is over the discrete and/or 
continuous parts of the spectrum.

In the case of the $1d$ Morse potential, Eq.(32) can be computed in a closed form
\bea
D_E(x,y| x^{\prime}, y^{\prime})
\sim({\cosh {d\over 2}})^{2i\nu-1}
{\Gamma({1\over 2}+\lambda-i\nu)\Gamma({1\over 2}-\lambda-i\nu)\over \Gamma(1-2i\nu)}\nonumber\\
F({1\over 2}+b-i\nu,{1\over 2}-\lambda-i\nu,1-2i\nu,{1\over \cosh^2d/2})~~~~~~~~~~~
\eea
where $E=\nu^2+{1\over 4}+\lambda^2$ and $F$ is the hypergeometric function.

Fourier transforming (33) one obtains a fundamental solution for the Morse potential 
\bea
d_{E,k}(x,x^{\prime})=\int dy e^{ik(y-y^{\prime})}D_E(x,y| x^{\prime}, y^{\prime})\sim ~~~~~~~~ \\
{\Gamma({1\over 2}-\lambda-i\nu)\over k\Gamma(1-2i\nu){\sqrt {x_{<}x_{>}}} }
M_{\lambda,-i{\sqrt {E}}}(2k{x_{<}})
W_{\lambda,i{\sqrt {E}}}(2kx_{>}) \nonumber
\eea    
In the zero field limit, Eqs.(33) and (34) reduce to 
\be
D_E(x,y| x^{\prime}, y^{\prime})\sim Q_{-1/2-i\nu}(\cosh d)
\ee
and
\be
d_{E,k}(x|x^{\prime})\sim I_{-i\nu}(kx_{<})K_{i\nu}(kx_{>})
\ee
where $x_{>/<}$ is the larger/smaller value of $x$, respectively.

Also, in the flat space limit, $kl\to\infty$, Eq.(33) reproduces the well-known result
\be
D_E(x,y| x^{\prime}, y^{\prime})
\sim{\Gamma({1/2-E/B})\over {\sqrt {Br^2}}}W_{E/B,0}(Br^2)
\ee
for the energies $E_n=B(2n+1)$ corresponding to the degenerate Landau levels, 
as all the scattering states are pushed to infinity. 

Another important calculable 
is the thermodynamic propagator ('heat kernal')
\bea
K_{\beta}(x,y|x^{\prime},y^{\prime})=<x,y|e^{-\beta H}|x^{\prime},y^{\prime}>=\nonumber\\
=\int dk\sum_{n/\nu} e^{ik(y-y^{\prime})-\beta E_{k,n/\nu}}
\psi_{k,\nu}(x)\psi_{k,\nu}(x^{\prime})
\eea
At zero field (i.e., in the case of the Liouville potential) it simplifies to 
\be
K_{\beta}(x,y|x^{\prime},y^{\prime})\sim
\exp(-{r^2\over \beta}-{\beta\over l^2})
{\sqrt {r/l\over \beta^2\sinh r/l}}
\ee
and can be used for studying the system's thermodynamic properties.
\\

{\it Thermodynamics and chaos}\\

A partition function for the (generalized) SYK action given by Eqns.(8) and (13) is 
represented by the field-theoretical path integral  
\be 
Z_{SYK}(\beta)=\int d{\phi}
\int^{\phi(\beta)=\phi}_{\phi(0)=\phi}D\phi(\tau)
e^{-\int_\tau S_{SYK}[\phi(\tau)]}
\ee
Alternatively, it can be computed in terms of the eigen-functions/values 
$\psi_{n/\nu}(\phi)$
and $E_{n/\nu}$ of the corresponding $1d$ Schroedinger equation
\be
Z_{SYK}(\beta)
=\int{d\phi}\sum_{n/\nu}|\psi_{n/\nu}(\phi)|^2e^{-\beta E_{n/\nu}}
\ee
On the other hand, the PMF partition function 
is represented by the world-line path integral  
\be
Z_{PMF}(\beta)=\int dxdy
\int^{x,y}_{x,y}Dx(\tau)Dy(\tau)e^{-\int_\tau S_{PMF}[x(\tau),y(\tau)]}
\ee
where $S_{PMF}$ is constructed from the same Hamiltonian (28). 
With the use of the eigenfunctions 
$\Psi_{k,n/\nu}(x,y)=\psi_{k,n/\nu}(x)e^{iky}$ it  
can be cast in the form similar to (42) 
\be
Z_{PMF}(\beta)
=\int dxdy\int dk\sum_{n/\nu}|\Psi_{k,n/\nu}(x,y)|^2e^{-\beta E_{k,n/\nu}}
\ee
thus establishing some form of equivalence between the (generalized) SYK and PMF problems. 

Alternatively, instead of performing a direct spectral 
summation the partition function can be deduced from  
the density of states (DOS)
\be
Z(\beta)=\int_0^{\infty} dE \rho(E)e^{-\beta E}
\ee
In turn, the (many-body) DOS of the SYK-like system can be read off from its 
single-particle PMF counterpart (32) 
\be
\rho(E)={1\over 2\pi}Im D_E(x,y|x,y)
\ee
In the Morse case, using the exact resolvent Eq.(33) one obtains the DOS in a closed form \cite{comtet} 
\be
\rho_{M}(E)\sim{\sinh 2\pi{\sqrt E}\over {\cosh 2\pi{\sqrt E}+\cos 2\pi\lambda}}
\ee
For $\lambda=0$, one then finds 
the well-known low-energy behavior the DOS in the SYK model 
$\rho(E)\sim{\sqrt E}$ \cite{syk,sy}.
In contrast, for $\lambda=1/2$ the DOS diverges as $\rho(E)\sim 1/{\sqrt E}$.
Notably, this behavior is reminiscent of that found in the SUSY version of the SYK model \cite{syk}.
On the other hand, a periodic dependence on $\lambda$ could be spurious
and remains to be better understood.

For $\lambda=0$, by performing an (inverse) Laplace transformation on (46) one can reproduce  
the low temperatures partition function of the Liouville model 
$Z_{L}(\beta)\sim \exp(O(l^2/\beta))/\beta^{3/2}$
for $\beta\gg 1/J$, while for $\beta\ll 1/J$
it yields $Z_L(\beta)\sim \exp(O(l^2/\beta))/\beta$. 
Thus, specific heat defined as $C=\beta^2\partial^2_{\beta}\ln Z(\beta)$
decreases with increasing temperature from $C=3/2$ down to $C=1$.

In contrast, the thermodynamic properties 
of the Morse model appear to be markedly different. Namely, for 
$\lambda=1/2$ specific heat rises from $C=1/2$ for $\beta J\gg 1$ to $C=1$
for $\beta J\ll 1$. 
Together with the aforementioned behavior of the density of states
this might be suggestive of a phase transition at $\lambda_c=1/2$. 

Such a conductor-to-insulator transition 
in the $SYK$ double-dot system has been studied, both, without \cite{qdot}
and with \cite{dvk2} such a realistic factor as Coulomb blockade taken into account. 
Conceivably, a bulk counterpart of this   
transition has long been 
predicted to occur between the SYK non-Fermi liquid and disordered Fermi liquid 
in a granular array of randomly $SYK_2$-coupled $SYK_4$ clusters \cite{24}. 

A difference between the two phases on the opposite sides of this purported 
transition can elucidated with the use of out-of-time-order correlators (OTOC).   
Generically, the OTOC amplitudes are expected to  
demonstrate some initial short-time/high temperature exponential growth 
\be
{<G_f(\tau_1,\tau_3)G_f(\tau_2,\tau_4)>\over <G_f(\beta/2,0)>^2}= 1-O({\beta J\over N})e^{\lambda_Lt}
\ee
revealed by summing the 'causal' ladder series and controlled by the 
chaotic Lyapunov exponent $\lambda_L$ \cite{syk}. 

The latter can be deduced directly from Eq.(11) for a general potential $V_{a,b}$ upon 
restoring a dependence of the fluctuating normal mode 
 $\delta g(\tau,T)\sim e^{\lambda_LT}\chi(\tau)$ on the 'center-of-mass' time $T$ 
and then continuing it analytically, $\tau\to it+\beta/2$  \cite{sy,syk}.

This way, one arrives at the eigenvalue equation in terms of the variable
$u=\tau/\beta$
\be
(\partial^2_{u}-W(g_0(u\beta)))
\chi=({\lambda_L\beta\over 2\pi})^2\chi
\ee
where $W(g(\tau))$ was defined after Eq.(11).

In the case of the Morse potential, one obtains the equation 
\be
\partial_u^2\chi+({\cos\theta\over \cosh u+\cos\theta}+
{2\sin^2\theta\over ({\cosh u+\cos\theta})^2})\chi=({\lambda_L\beta\over 2\pi})^2\chi
\ee
where the effective potential crosses over from $W_q=-2/\cosh^2u$ in the pure $SYK_q$ limit
($\theta\to\pi/2$) with the ground state $\chi_q\sim 1/\cosh u$ to $W_{q/2}=-1/2\cosh^2(u/2)$ in the pure $SYK_{q/2}$ one ($\theta\to 0$).  
In both limits, the Lyapunov exponent 
approaches its maximal value  $\lambda^{max}_L=2\pi/\beta$ \cite{syk} 
as $\lambda_L/\lambda^{max}_L=1-O(max[1/\beta J, \Gamma^2\beta/J])$
for $\Gamma\ll J$ and $1/J\lesssim\beta\lesssim J/\Gamma^2$
or $=1-O(J/\Gamma^2\beta)$ for $\Gamma\gg J$ and $J/\Gamma^2\lesssim\beta\lesssim 1/\Gamma$ \cite{hybrid}. 

In the special case of $q=4$, though, the fixed-point 
$SYK_2$ behavior corresponds to the disordered but non-chaotic 
Fermi liquid where $\lambda_L$ is expected to vanish. 

In the intermediate regime and for $q>4$ the Lyapunov exponent appears
to take lower, yet non-zero, values \cite{dvk4}. 
It does not vanish at any finite temperature, though, thus calling for a closer look at  
any scenario of a genuine finite-temperature phase transition - or a zero-temperature 
one predicted to occur at a critical ratio $\Gamma/J$ vanishing at large $N$ as a power of $1/N$
\cite{bak,24}.   
In that regard, it would be particularly interesting to compute $\lambda_L$ at the super-symmetric point $J=2{\Gamma}$.

Also, in the aforementioned 'variable scaling' model \cite{hybrid} 
some non-maximal and non-universal, yet temperature-independent and growing with 
the increasing integer parameter $n$,  
values of $\lambda_L$ were reported on the basis 
of the numerical solution of (48).
In turn, the Hulten potential falls somewhere in between 
the 'super-symmetric' point at the $SYK_q-SYK_{q/2}$ model
and the 'variable scaling' one \cite{dvk4}. 

As an interesting consistency check, the eigenfunction equation for the 
Toda potential $V_{2,-2}$ evaluated on the solution (26) 
satisfies the same Eq.(49) apart from the constant shift,  
$W_T= -{2/\cosh^2u}+1$, which raises the ground state energy to zero, thus implying 
$\lambda_L=0$. This observation is consistent with the non-chaotic
nature of the discrete spectrum consisting only of the bound states.
\\

{\it Dual gravities}\\

In $2d$, a powerful gauge invariance under 
local coordinate diffeomorphisms eliminates any bulk degrees of freedom, thereby 
making such theories locally quantum trivial in the absence of matter.
Such bulk theories appear to be topological and allow for explicit classical solutions,
thereby providing natural candidates for testing out the foundations of the holographic principle. 

Moreover, the gauge symmetry leaves only one independent metric component
(e.g., $g_{01}=g_{10}=0,g_{00}=1/g_{11}$), thus reducing (up to a conformal factor)
all the static (Euclidean) metrics to the set $ds^2=e^{2\nu(x)}d\tau^2+e^{-2\nu(x)}dx^2$ parametrized by a single  function $\nu(x)$.

However, a $2d$ gravity theory can still develop a non-trivial 
boundary behavior as a result of introducing either an additional dilaton, Liouville, or scalar matter field. Alternatively, it requires anisotropic space vs time scaling,
$\tau\sim x^z$, characterized by a dynamical critical index $z$. 
Thus, such extensions can be sought-out not only 
in the context of generalized JT but also the Horava-Lifshitz (HL) \cite{hl} theories.   

The original (ostensibly $2d$) 
JT model is well known to be described by the Schwarzian boundary action (8) 
providing a natural holographic connection to the edge 
modes propagating along the $1d$ boundary  \cite{jt,syk}. Indeed, the Schwarzian can be directly 
related to the extrinsic curvature of a fluctuating closed $1d$ boundary of a $2d$ region,  
$
K=1+Sch{\{}\tan{\pi f/\beta},\tau {\}}+\dots
$.

In practical terms, establishing generalized holographic duality 
with a given Liouville-type theory described by Eqns.(8) and (13) 
can be formulated as a task of constructing the  
bulk theory whose boundary dynamics is governed by the same $1d$ Hamiltonian
as that of the conjectured dual $1d$ quantum system.

In that regard, the boundary actions of SUSY 
and higher spin extensions of the $2d$ dilaton gravities  
were argued to represent certain specific limits of the generalized JT model,
including its non- and ultra-relativistic variants.
Alternatively, the complex SYK model was argued to have a possible 
flat space bulk dual \cite{jt}.  

The most general action incorporating a dynamical dilaton \cite{jt} 
\be
S_d=\int dxd\tau{\sqrt g}(RU(\Phi)+V(\Phi)(\partial\Phi)^2+W(\Phi))
\ee
is parametrized in terms of the functions $U$, $V$, and $W$ of the dilaton field $\Phi$.
Such generalized dilaton gravities have been encountered among the 
deformations and compactifications of the higher-dimensional theories.
Among them, there is an important family of potentials 
$U\sim V\sim W\sim\Phi$ which may allow for the $AdS_2$ ground state.

Moreover, the so-called $F(R)$-gravities with the generic 
action $S_{F(R)}=\int dxd\tau F(R)$ were argued to be all equivalent to the 'minimal' 
JT action with $F(R)=\Phi R-V(\Phi)$ with the expectation value of the dilaton field 
being related to the Ricci curvature, as per the equation $R=\partial V/\partial\Phi$ \cite{F(R)}.

Alternatively, the intrinsically topological nature of the JT gravity  
can be made manifest with the use of its $1st$ order formulation \cite{jt2} 
\bea
S_{JT}=\int (\Phi \epsilon_{\mu\nu}d^{\mu}\omega^{\nu}
+W(\Phi)\epsilon^{ab}\epsilon_{\mu\nu}e^{\mu}_a e^{\nu}_b+\nonumber\\
X^a\epsilon_{\mu\nu}d^{\nu}e^{\mu}_a+X^a\epsilon^b_a\omega_{\mu} e^{\mu}_b)
\eea
in terms of the vielbein $e^{\mu}_a$ and spin-connection $\omega^{\mu}$ 
which is independent of the background metric 
$g_{\mu\nu}=\eta_{ab}e^a_{\mu}e^b_{\nu}$ (here $\eta_{ab}$ is the flat space metric). 
Notably, the action (51) shares its topological nature with the  
$3d$ gravity that can be cast in terms of the (twinned) Chern–Simons theory \cite{cs}.

Another viable candidate to the role of a 
$2d$ gravity dual to a generalized SYK model 
can be sought out in the form of the (Lorentz non-invariant) Horava-Lifshitz action \cite{hl}
\be
S_{HL}=\int dxd\tau {\sqrt {g}}N(aK^2+b\Lambda+c(N^{\prime}/N)^2)
\ee
where $a, b, c$ are numerical parameters, $\Lambda$ is the $2d$ cosmological term,
$N$ and $N_1$ are the lapse and shift functions,
$h={\sqrt g_{xx}}$, and $K=-({\dot h}/h-N_x^{\prime}/h^2+N_xh^{\prime}/h^3)/N$,
the dots and primes standing for the time and space derivatives, respectively.
In contrast to the dilaton gravity (51)
Eq.(53) is only invariant under the foliation-preserving 
diffeomorphisms $\tau\to \tau^{\prime}(\tau)$ and $x\to x^{\prime}(x,\tau)$.

Certain previously proposed $F(R)$-HL theories 
provide the Lifshitz-type black hole solutions
with a constant negative curvature $R=-2z^2/l^2$. 

Under the projectability condition \cite{hl} 
one can choose $N=N(\tau)$ to be a global (spatially uniform)
variable which then gives rise to $K^{\prime}=0$ (hence, $K=K(\tau)$).
Furthermore, by using the coordinate gauge symmetry one can fix $N=1$ and $N_x=0$.  
The number of primary and secondary Hamiltonian constraints then equals 
the dimension of the phase space, thus 
reducing the number of the dynamical bulk degrees of freedom to zero.
From that one incurs that the conjugate 
momentum $p=p(\tau)$ is independent of the spatial position either.
  
Canonically quantizing the action (52) one then arrives at the effectively $1d$ PMF-like Hamiltonian 
\be
H_{HL}=aqp^2+b\Lambda q+c{P\over q^w}
\ee 
where $q(\tau)=\int dx h(x,\tau)$ and $p(\tau)$ constitute a pair of conjugate canonical variables 
while the $2nd$ variable $Q(\tau)$ is cyclic and paired up with a conserved conjugate momentum
$P({\tau})=P$. 

In the context of Friedmann-Robertson-Walker (FRW) cosmology, 
the Hamiltonian (53) emerges in the Wheeler-DeWitt equation, the parameter $w$ taking values $1,0,-1$ 
for radiation, matter, and dark energy, respectively. 
Contrasting Eq.(53) against Eq.(31) one finds that the essential terms in the 
two expressions match for, e.g., $\alpha=1/2,\beta=-w-1/2, \gamma=-1-w$.
 
Adding matter to (53) introduces another pair of conjugate variables, similar to 
the formulation of the PMF problem in a non-separable gauge.   
In particular, the projectable HL action with an additional scalar field $Q(x,\tau)=Q(\tau)$
governed by a potential $V(Q)$ and paired with a 
conjugate momentum $P(\tau)$ produces the Hamiltonian which
resembles the $2d$ PMF Hamiltonian 
\be
H_{HL+S}=aqp^2+b\Lambda q+c{P^2\over q}+qV(Q)
\ee 
\\

{\it Summary}\\

The orthodox holographic scenario requires a bulk gravity  
to have non-trivial dynamics that gets quenched and turns 
classical only in a certain ('large-$N$') limit \cite{ads}. 

In that regard, the SYK-JT duality would often be referred to as the case of 'bona fide'
low-dimensional holographic correspondence. It is generally agreed, 
though, that such equivalence does not quite 
rise to the level of the full-fledged AdS/CMT holographic duality,  as 
the JT bulk dual is non-dynamical and determined 
by the boundary degrees of freedom, thus making both systems effectively $1d$.

This note argues that similar (pseudo)holographic 
relationships can be established between the various extensions 
of the original SYK model and more general (JT, $F(R)$-, HL, etc.) $2d$ gravities.
The correspondence between their low-energy sectors  
presents a form of equivalence between different  
realizations of the co-adjoint orbits of the (chiral) Virasoro group.

Formally, both sides of such duality can be described in terms of some $1d$ Liouvillean quantum mechanics, thus generalizing the pure Schwarzian action which description can also be mapped onto an equivalent 
(single particle) PMF problem. From the practical standpoint, certain analytically solvable quantum-mechanical potentials can then be related to the physically relevant SYK deformations, such as 
the action given by Eqns.(8) and (14) for a double SYK quantum dot.   

The PMF analog picture allows for direct access to the resolvent $D_E$
and heat kernel $K_{\beta}$ functions, thus allowing one to compute the density of states $\rho(E)$, 
partition function $Z(\beta)$, and other thermodynamic properties
of the boundary SYK-like system of interest.
By further utilizing this approach one can also study the various
quantifies of entanglement, quantum chaos, and even more subtle $n\geq 2$-body correlations. 

Furthermore, the tangle of (pseudo)holographic 
relationships between the $SL(2,R)$-symmetric boundary (Schwarzian/Liouville-like) and bulk (JT/HL-like) models can be viewed as different forms of embedding (at fixed radial and angular vs temporal and angular coordinates, respectively) into the global $AdS_3$ space \cite{ads3}. 
Importantly, a similar relationship also exists between the $1+2$-dimensional gravity with its  Banados–Teitelboim–Zanelli black hole backgrounds and the various (e.g., Korteweg-de-Vries) families of solvable $1+1$-dimensional quantum systems \cite{cs}. Among other things, such equivalence can be utilized to study non-linear
hydrodynamics of the soliton-like edge states of generalized bulk QHE systems \cite{hydro}.
 
Thus, when seeking out genuine implementations of the central holographic 'IT from QUBIT' paradigm one 
might first want to make sure that the conjectured duality does not appear to be
of the 'ALL from HALL' variety.
Indeed, discovering a (possibly, hidden) topological origin of holographic 
correspondence could greatly help to demystify this otherwise fascinating, yet baffling, concept.   
\\

\end{document}